\documentstyle[prl,aps,multicol,epsf]{revtex}
\begin{document}
\draft
\title{Theory of Hysteresis Loop in Ferromagnets.}
\author{Igor F.Lyuksyutov$^{a,*}$, Thomas Nattermann $^{b,d}$,
and Valery Pokrovsky $^{a,c}$}
\address{
$^{a}$ 
Department of Physics, Texas A\&M University,\\
College Station, TX 77843-4242, U.S.A.\\
$^b$ Institut f\"ur Theoretische Physik,\\
Universit\"at zu K\"oln, 50937, K\"oln, Germany.\\
$^c$ Landau Institute for Theoretical Physics, Moscow, Russia\\
$^d$ Ecole Normale Superieure
Laboratoire de Physique Theorique\\
24,rue Lhomond
75231 Paris Cedex 05, France.}

\date{\today}
\maketitle
\begin{abstract}
We consider three mechanisms of hysteresis phenomena in 
alternating magnetic field:
the domain wall motion in a random medium, the nucleation and
the retardation of magnetization due to slow (critical) 
fluctuations. We construct quantitative theory for all
these processes. The hysteresis is characterized by two dynamic
threshold fields, by coercive field and by the so-called
reversal field. Their ratios to the static threshold field
is shown to be function of two dimensionless variables
constituted from the frequency and amplitude of the ac field as well
as from some characteristics of the magnet. The area and the shape
of the hysteresis loop are found. We consider different
limiting cases in which power dependencies are valid.
Numerical simulations show the domain wall formation and
propagation and confirm the main theoretical predictions.
Theory is compared with available experimental data.
\end{abstract}
\pacs{75.70.Ak,75.60.Ej, 75.60.Ch}
\begin{multicols}{2}
\section
{Introduction}
\label{introduction}
The hysteresis loop (HL) has been first studied more
than century ago  \cite{stein}. 
However, the understanding of this
process in  thin magnetic films as well as in bulk magnets is still 
rather poor. 
Many efforts have been devoted recently to prediction 
(see \cite{pelc,dhar,acharyya,zang,madan}) 
and experimental verification (see \cite{wang1,wang2}, 
\cite{jim}) of
the scaling behavior of the hysteresis loop area (HLA)
as a function of the applied magnetic field 
frequency and  amplitude for
thin magnetic films (for a  brief review of HLA 
scaling results see \cite{jim}).
The scaling behavior of the HLA have been first reported 
in the pioneer work \cite{stein} for 3D magnets.
While there exists an extended literature on
the hysteresis of 3d magnets, the properties of
HL in 2d systems are much less known. There are only few
articles devoted to the HL in ultrathin ferromagnetic
films \cite{bruno}, \cite{raquet}, \cite{wang1,wang2},
\cite{jim}, 
though the hysteresis effects have
been found as a side effect in many others
(see, for example \cite{erskine}, \cite{beck}). 
Critical exponents found in the experiments
with thin films 
vary dramatically for  different materials
(see e.g. \cite{wang1,wang2} and \cite{jim})
and probably for different regimes. Different authors
disagree each to other (see the already cited
articles \cite{wang1,wang2} and \cite{jim})
and also disagree with numerical
simulations \cite{pelc}.

Several years ago
mean field type models with 
single \cite{bruno} or many 
\cite{raquet} relaxation times have been applied to 
analyze experimental data.
The authors of  \cite{bruno,raquet}
assumed that the
HL was  controlled by the nucleation process.
These authors predicted the logarithmic dependence
of the coercive field $h_c$ on the  rate of the  applied
magnetic field $\dot h$.  
In a recent experiment \cite{jim} it was found  that  the HLA 
depends on frequency of applied field 
as a power with a small exponent
($\sim 0.03 - 0.06$) or, possibly, logarithmically.
However, in the framework of
the same approach the HLA
must behave also as logarithm of $h$ 
with the same coefficient (exponent). This dependence
has never been observed in the experiment.
Therefore, we propose a new analysis of such a HL in this
article.

The purpose of this article is to formulate a rather general
approach to magnetization reversal mechanisms and to indicate 
several important measurable characteristics of the HL besides of the HLA.
We will see that these characteristics 
are governed by two dimensionless parameters combined
from the field frequency $\omega$, its amplitude $h_0$
and characteristics of the magnetic material.
Everywhere in what follows we assume that the external
field varies harmonically in time $h(t)=h_0\sin\omega t$.

The hysteresis behavior  may have various origin.
It can be mediated by nucleation process,
by the DW propagation or simply
by retardation of the magnetization
due to fluctuations. We consider all these 
mechanisms and establish conditions at which
one of them is dominant. 

Defects play an important role in the DW propagation.
They create a finite threshold value $h_p$ of static magnetic field
necessary for the DW depinning. The threshold field $h_{t1}$ in
the dynamical problem can differ substantially from $h_p$.  
We find that in a medium
with defects the moving DW, passing rare extended
defects, may form
bubbles of reversed spins. These bubbles play an important role as
prepared nuclei in the next half-cycle of the magnetization reversal. 

In this article we consider magnets of the Ising (uniaxial) symmetry.
Their properties may be very different depending on the
strength of the anisotropy. In the experimentally 
studied films the anisotropy was very weak. In this case
the domain wall width is large in comparison to the
lattice constant. On the contrary, in the original Ising
model the anisotropy is assumed to be large and DW width $l$
is simply the lattice constant.
However, these different models become equivalent after 
a simple rescaling:
the DW width should be accepted as a new elementary (cut-off)
length. It means that we consider a spin cluster of the linear size  $l$
as a new elementary spin. This approach allows us to apply
the Ising model supplied with the  Glauber dynamics for numerical
simulations. 

Peculiarities of the two-dimensional situation are:  much higher
mobility of the DW as well as much stronger fluctuations.
This makes the experimental situations as well as the theoretical
description much more diverse than those for a 3D magnet.

This article is organized as follows. In sections \ref{dw} and \ref{line} we
consider the individual DW motion. Equation of motion is 
formulated and justified in Sect.\ref{dw}. It is solved in Sect.\ref{line}.
In the same section we introduce characteristic fields
$h_{t1}, h_{t2}, h_c, h_r$, the HLA $\cal A$, find the scaling arguments
and analyze several limiting regimes in which simple power
scaling is valid. In Sect.\ref{traps} the process of the bubble formation
is studied. Sect.\ref{nucleation} is devoted to the HL controlled by the
nucleation process. The HL driven by strong magnetization 
fluctuations, especially
near the Curie point, is considered in Sect.\ref{adiabatic}. Numerical 
simulations of the HL and the domain structures for the
2d Ising model with Glauber dynamics which supports
results obtained in Sect.\ref{line}, Sect.\ref{traps} 
are presented in Sect.\ref{mc}. 
In Sect.\ref{conclusions} we summarize our results and compare 
them to the experimental data.
In the rest of this article we use the notation $h$ for
magnetic field, $m$ for magnetization and $ M$ for
the total magnetic moment of a magnet.

\section {Domain wall motion in a random medium}
\label{dw} 
As we already mentioned in the Introduction, our starting point is 
an impure 
ferromagnet with either weak or strong Ising anisotropy.
The soft spin version of the system is then given by a $\phi^4$
model with a bare domain wall width $l \gg a$, where $a$ 
denotes the original lattice spacing.
The imperfections in the model may be in principle either of 
random bond (i.e. random $T_c$) or 
random field type. We will argue below, that in the 
region we are mainly interested in, namely above the depinning threshold,
both types of impurities act essentially as random field impurities. 

As it was shown by Bausch et al. \cite{Bausch} 
(see also \cite{feigel}\cite{koplik}), 
equation of motion for a domain
wall without overhangs 
can be written in a following way
\begin{equation}
\frac{1}{\gamma\sqrt{g}}\frac{\partial {\it Z}}{\partial t}=
\Gamma {\bf\nabla}\cdot
\left( g^{-1/2}{\bf\nabla}{\it Z}\right) + h + \eta ({\bf x},{\it Z})
\label{eq:1}
\end{equation} 
where ${\it Z} ({\bf x},t)$ denotes the interface 
position and  $g=1+({\bf\nabla} {\it Z} )^2$. $\gamma$ and $\Gamma$ 
are the  domain wall mobility 
and stiffness, respectively. 
 $h=\mu_BH{ M}$, where $H$
is the external magnetic field and ${ M}$ 
is the magnetization.
Finally $\eta$ denotes the random force generated by the impurities.

For broad domain walls $\Gamma\approx J/(a^{D-1}l)$, where $D$ denotes 
the dimensionality of the wall. For narrow walls $\Gamma$
depends in general on $J,\;T$, and the disorder strength in a complicated 
way \cite{nat82} . 

The  random fields  
 $\eta (\vec{r}=({\bf x},{\it Z}))$ generated 
by imperfections is 
assumed to be Gaussian distributed and short range correlated 
with $\overline{\eta (\vec{r})}=0$ and

\begin{equation}
\overline{\eta ({\vec r})\eta ({\vec r}^{\,\prime})}=\eta^2 l^{D+1}
\delta_l(\vec r-{\vec r}^{\,\prime}).
\label{eq:2}
\end{equation}
Here $\delta_l(\vec r)$ denotes a delta-function smeared out over a 
distance $l$. As was first argued by Narayan and Fisher \cite{fn93},
in the region above the depinning threshold random bond and 
random field impurities act in a similar way. This can be seen most easily
from an example of two ratcheted potential (see figure 1), one for random 
bond and another for random field impurities. Although the potential 
$V({\it Z})=\int_0^{\it Z} \eta ({\bf x},{\it Z}')\,d{\it Z}'$ 
in the random field case (figure 1b) 
has fluctuations which scale like ${\it Z}^{1/2}$, 
it leads to the same random forces 
as the random field potential (figure 1a) which shows of order one 
fluctuations. Therefore, we restrict ourselves throughout this paper 
to the case of random field impurities.
Moreover, we will assume that the disorder is
weak, i.e. that the condition
\begin{equation}
\Gamma\gg\eta l
\label{eq:3}
\end{equation}
is fulfilled.

For $|{\bf\nabla}{{\it {\it Z}}} |^2\ll 1$, 
$ g\approx 1$ and the equation of motion 
takes the form considered previously \cite{feigel} \cite{koplik} 
\cite{nat92} \cite{fn93}. Below
we summarize some of the results found in \cite{nat92}\cite{fn93}: 
Since the disorder is weak (see (\ref{eq:3})), the interface is 
essentially flat on length scales $L\ll L_c$ where
\begin{equation}
L_c\approx l\left(\frac{\Gamma}{\eta l}\right)^{2/(4-D)}\gg l\;.
\label{eq:4}
\end{equation}
is the so-called Larkin length. 
On larger scales the wall can adapt to the disorder and, as a result, it gets 
pinned for driving fields $h\stackrel{<}{\sim}h_p$ with
\begin{equation}
h_p\approx\gamma lL_c^{-2}=\eta\left(\frac{\eta l}{\Gamma}\right)^{D/(4-D)}
\ll\eta
\label{eq:5}
\end{equation}
for the pinning threshold.
If $h$ exceeds $h_p$, the wall starts to move. For $h\gg h_p$ the influence 
of the disorder is weak and the velocity is proportional to the
driving field
\begin{equation}
v=\left<\dot{{\it D}} \right>\approx\gamma h
\label{eq:6}
\end{equation}
Corrections to this relation can be considered in the framework of 
high-velocity expansion which can be expressed as a power series in
\begin{equation}
\frac{\xi_v}{L_c}\approx\left(\frac{h_p\gamma}{v}\right)^{1/(z -\zeta )}\,,
\label{eq:7}
\end{equation}
where $\xi_v$ is dynamical correlation length which diverges as 
$v\rightarrow 0$; $z$ and $\zeta$ are the dynamical and the roughness 
exponent, respectively. Outside the dynamical critical region, i.e., for 
$h-h_p\gg h_p$, $z =2$ and $\zeta =0$, respectively.

If the driving field $h$ is so small,
that $\xi_v\stackrel{>}{\sim}L_c$,
the high-velocity expansion breaks down and a (functional) renormalization 
group calculation has to be applied   \cite{nat92}. This leads to a 
renormalization of the
mobility constant $\gamma\rightarrow\gamma_{\rm eff}$ with
\begin{equation}
\gamma_{\rm eff}\approx\gamma\left(\frac{\xi_v}{L_c}
\right)^{(1/3)(4-D-\zeta )}\;.
\label{eq:8}
\end{equation}
After integrating out the interface fluctuations on the length scales
$L\stackrel{<}{\sim}\xi_v$, the effective equation of motion for the 
interface profile 
${\it Z}({\bf x},t)=\left< {\it Z} ({\bf x},t)\right>_{\xi_v,t_v} $
on large scales is given by
\begin{equation}
\frac{1}{\gamma_{\rm eff}}\frac{\partial {\it Z}}{\partial t}=
\Gamma {\bf \nabla}^2{\it Z} + h - h_p+\eta_{\rm eff}({\bf x},vt)\;.
\label{eq:interface profile}
\end{equation}
Here $\left<\right>_{\xi_v,t_v}$ denotes the spatial and
time average over scales $\xi_v$ and $t_v$, respectively and $\eta_{\rm eff}$
is the  renormalized random field which acts as a thermal noise. Since the 
latter leads to an interface roughness characterized by the exponent
$\zeta_0 =(2-D)/2$, we may neglect the influence of the random field
on these length scales. The mean velocity of the interface is given by
\begin{equation}
v\approx\gamma h_p\left(\frac{h-h_p}{h_p}
\right)^{\theta}\,,\quad \frac{h-h_p}{h_p}\ll 1
\label{eq:9}
\end{equation}
where $\theta =({z} -\zeta )/(2-\zeta )$. 
$z$ and $\zeta$ take now non-trivial values,
which can be calculated in by $\epsilon$-expansion in $D=4-\epsilon$ or determined
numerically. For $D=1$ the $\epsilon$-expansion gives $\zeta =1$ and $z=4/3$
and hence $\theta=1/3$ whereas the numerical values are $\zeta =5/4$,
$z\approx 1.42$ and $\theta\approx 1/4$ \cite{lesch}. To unify our results for the domain
wall velocity we rewrite (\ref{eq:6}), (\ref{eq:9}) as
\begin{equation}
v\approx h_p\gamma f\left(\frac{h}{h_p}-1\right)\,,
\quad f(x)\approx\left\{\begin{array}{l@{\;,\;}l}
x^{\theta} & x\ll 1\\ x & x\gg 1\end{array}\right.
\label{eq:10}
\end{equation} 

On the length scales $L_c\ll L\ll \xi_v$ the domain wall is rough
\begin{equation}
w(L)=\left<\left( {\it Z} ({\bf x}_1)-{\it Z} ({\bf x}_2)\right)^2
\right>^{1/2}_{|{\bf x}_1-{\bf x}_2|=1}\approx l(L/L_c)^{\zeta}\;.
\label{eq:11}
\end{equation}
Bumps in the domain walls which emerge from random clusters heal on time
scales
\begin{equation}
t_v\approx\frac{l}{v}\left(\frac{h_p}{v\lambda}\right)^{\zeta /(z -\zeta)}
\approx\frac{l}{\gamma h_p}\left(\frac{h}{h_p}-1\right)^{-z /(2-\zeta )}
\label{eq:12}
\end{equation}

However, on larger scales $L\gg\xi_v$ the random field acts merely as 
a thermal noise and the roughness exponent is reduced to $\zeta =1/2$ and
$\zeta =0$ (log) in $D=1$ and $D=2$ dimensions, respectively.

\section{Motion of a rectilinear domain wall.}
\label{line}
We start with description of a rectilinear domain wall motion. 
In previous section we demonstrated that the domain wall roughness
can be ignored on a time scale $t>t_v$. Thus, locally the domain
wall moves as a straight line. In some experiments only one
domain wall survives (see sect. 7). In this case the model problem
of rectilinear domain wall motion is close to reality. 
In other cases this problem is an important part of more
complex problem describing either local properties of domain
wall motion or the order of magnitudes for the global motion.
Thus, we consider the motion of a rectilinear domain wall
under the action of magnetic field antiparallel to
the magnetization.
 
We have mentioned already that the fluctuation bending of the domain wall
can be neglected if the
characteristic time of the process is much 
more than the bump healing time
$t_v\propto v^{-z/(z-\zeta )}$ . This requirement
suggests that $\omega t_v\ll 1$. Anyway, this requirement
must be satisfied since, otherwise the average position of the domain
wall almost does not change during half a period of oscillations.

The domain wall is assumed to be fixed at the left boundary of the
sample $Z=0$ at the initial moment.
We will solve equation of motion (\ref{eq:6}) for domain wall coordinate
for harmonically oscillating magnetic field $h=h_0\sin{\omega t}$. 
Instead of integrating it over time, we integrate it over field
by a following change of coordinates:
\begin{equation}
dt\,=\,{1\over \omega}\frac{dh}{\sqrt{h_0^2-h^2}}
\label{time-field}
\end{equation}
After integration we find an expression for $Z$ vs. magnetic
field $h$:
\begin{equation}
Z\,=\,{\gamma\over\omega}\int_{h_p}^h f\left(\frac{h-h_p}{h_p}\right)
\frac{dh}{\sqrt{h_0^2-h^2}}.
\label{z(h)}
\end{equation} 
This equation is correct for $h>h_p$. For smaller value of $h$
the domain wall does not move: $Z=const$. Equation (\ref{z(h)})
should be complemented by a prescription to change the sign of
the square root each time as $h$ reaches its maximum or minimum
value $\pm h_0$ and by an initial condition $Z=0$ at $t=h=0$.
The second necessary prescription is to substitute $h-h_p$ by
$-h-h_p$ when $h$ is negative.
To transfer from the coordinate $Z$ to the magnetic moment $\cal M$,
a following representation is useful:
\begin{equation}
{ M}\,=\,{M}_s{2Z-L\over L}
\label{magmoment}
\end{equation}
where $M_s$ is the saturation magnetic moment.
First of all we find two important boundary values for the amplitude $h_0$
which separate hysteresis loops of different shapes. The first of them
is the dynamic threshold field $h_{t1}$, the smallest value of $h_0$ at which 
the domain walls reaches
the right boundary of the sample $Z=L$. At $h_0<h_{t1}$ the magnetization
is not reversed fully and the hysteresis loop is asymmetric 
(see Fig \ref{fig:loop}a). At larger values of $h_0$ the hysteresis loop is 
symmetric 
under inversion $h\longrightarrow -h, { M}\longrightarrow -{ M}$
(Fig. \ref{fig:loop}c). The value of $h_{t1}$ is determined by a following 
equation:
\begin{equation} 
\frac{\omega L}{2\gamma h_p}\,=\,\int_1^{h_{t1}/h_p}f(x-1)
\frac{dx}{\sqrt{(h_{t1}/h_p)^2-x^2}}.
\label{h-t}
\end{equation}
Note, that at $h_0=h_{t1}$ the hysteresis loop is symmetric with respect
to reflections in the axis $h$ and $\cal M$ (Fig. \ref{fig:loop}b). 
From Eqn. (\ref{h-t}) 
it is seen that the ratio $h_{t1}/h_p$ is the function of one dimensionless
variable $\omega L/(\gamma h_p)$. This scaling relationship will be 
analyzed in details later.

The next notorious field is $h_{t2}$ defined as the value of $h_0$
at which the domain wall reaches the right end of the sample $Z=L$
during one fourth of period, just at $h=h_0$. Equation which defines
$h_{t2}$ is rather similar to that for $h_{t1}$:
\begin{equation} 
\frac{\omega L}{\gamma h_p}\,=\,\int_1^{h_{t2}/h_p}f(x-1)
\frac{dx}{\sqrt{(h_{t2}/h_p)^2-x^2}}.
\label{h-1}
\end{equation}
It differs from Eq. (\ref{h-t}) by the absence of factor 2 in denominator
of the l.-h. s. The hysteresis loops corresponding to $h_0>h_{t2}$ acquire
characteristic ``whiskers'', single-valued pieces of the curve 
$M(h)$ which are absent in hysteresis curves for $h_0<h_{t2}$
(see Fig. \ref{fig:loop}d).

At a fixed $h_0>h_{t1}$ it is possible to define the coercive field
$h_c$ by the requirement ${ M}(h_c)=0$. Finally, for $h_0>h_{t1}$
the so-called reversal field $h_r$ can be defined as a value of
the field $h$ at which the magnetic moment reverses fully.
At the value $h=h_r$
two branches of the hysteresis curve intersect each other.
In other words, $Z(h_r)=L$. At $h$ between $h_r$ and
$h_0$ the magnetic moment remains a constant ${ M}={ M}_s$.
The values $h_c$ and $h_r$ are shown in Fig. \ref{fig:loop}.  
Using equations (\ref{z(h)}) and (\ref{magmoment}), we find:
\begin{equation}
\frac{\omega L}{2\gamma h_p}\,=\,\int_{h_p}^{h_c}f\left(\frac{h-h_p}{h_p}
\right)\frac{dh}{\sqrt{h_0^2-h^2}}.
\label{h-c}
\end{equation}
Equation for $h_r$ reads:
\begin{equation}
\frac{\omega L}{\gamma h_p}\,=\,\int_{h_p}^{h_r}f\left(\frac{h-h_p}{h_p}
\right)\frac{dh}{\sqrt{h_0^2-h^2}};\,\,\,\,\,\,(h_0>h_{t2})
\label{h*1}
\end{equation}
\begin{eqnarray}
\frac{\omega L}{\gamma h_p}&=&
\left[\int_{h_p}^{h_r}+2\int_{h_r}^{h_0}\right]\\
&\times&f\left(\frac{h-h_p}{h_p}
\right)\frac{dh}{\sqrt{h_0^2-h^2}};
\,\,\,\,\,\,(h_{t1}<h_0<h_{t2}).
\label{h*2}
\end{eqnarray}
The ratios $h_c/h_p$ and $h_r/h_p$ are functions of 2 dimensionless
variables $u = \omega L/(\gamma h_p)$ and $v =h_0/h_p$. 
Note that by knowledge
of $w_2=h_{t2}/h_p$ (or $w_1=h_{t1}/h_p$) as a function of parameter 
$u=\omega L/\gamma h_p$
one can restore the function $f(x)$ solving the Abelian equation:
\begin{equation}
f(x-1)\,=\,{2\over\pi}{d\over dx}\int_1^x
\frac{u(w)wdw}{\sqrt{x^2-w^2}}.
\label{restore}
\end{equation}
The area of hysteresis loop $\cal A$ can be also expressed in the 
integral form:
\begin{equation}
{\cal A}\,=\,4h_p{M}_s\,-\,
4{{ M}_s\over L}\int_{h_p}^{h_r}Z(h)dh.
\label{area} 
\end{equation}
Now we proceed to the analysis of the hysteresis loop characteristics.
It is controlled by parameters $u=\omega L/\gamma h_p$
and $v=h_0/h_p$. We start with small $u\ll 1$ imposing no restrictions
on the value $v$. First we show that, at small $u$, the fields $h_{t1}$
and $h_{t2}$ are close to $h_p$. Indeed, it is clearly seen from eqns.
(\ref{h-t}) and (\ref{h-1}). Solving them approximately and employing
the asymptotic formula for $f(x)$ at small $x$ (\ref{eq:10}), we find
\begin{equation}
u/2\,=\,{1\over\sqrt{2}}\int_1^{h_{t1}/h_p}
(x-1)^{\theta}\frac{dx}{\sqrt{(h_{t1}/
h_p)-x}}.
\label{h-t-1}
\end{equation}
Introducing a new integration variable 
$s=\frac{x-1}{(h_{t1}/h_p)-1}$, eqn. (\ref{h-t-1})
can be transformed as follows:
\begin{equation}
u/2\,=\,{1\over\sqrt{2}}\left({h_{t1}\over h_p}-1\right)^{\theta +1/2}
B(\theta +1, 1/2),
\label{B}
\end{equation}
where $B(x, y)$ is the Euler beta-function. Using its standard
representation \cite{gradstein}, one finds:
\begin{equation}
{h_{t1}\over h_p}-1\approx\left[\frac{(1+1/2\theta)\Gamma(\theta +1/2)}
{\sqrt{2\pi}\Gamma(\theta )}u\right]^{{2\over 2\theta +1}}.
\label{h-t-2}
\end{equation}
In a similar way the asymptotic of the field $h_{t2}$ can be established.
The ratio $(h_{t2}-h_p)/(h_{t1}-h_p)$ does not depend on the parameter $u$
if $u$ is small:
\begin{equation}
\frac{h_{t2}-h_p}{h_{t1}-h_p}\,=\,2^{{2\over 2\theta+1}}.
\label{ratio}
\end{equation}
The shape of the hysteresis loop and its area depends not only on the
parameter $u$, but also on the parameter $v=h_0/h_p$. If $v$ is close to
1, i.e. if $h_0$ is also close to $h_p$, one can employ the low-field
asymptotic of the function $f(x)$ (\ref{eq:10}) also in equation (\ref{h*1})
for $h_r$. Then, for $h_0>h_{t2}$ approximate equation for $h_r$
reads:
\begin{equation}
u\,=\,{1\over\sqrt{2}}\int_1^{h_r/h_p}(x-1)^{\theta}\frac{dx}
{\sqrt{v-x}}.
\label{h*-1}
\end{equation}
Introducing a new integration variable $s={x-1\over v-1}$, we find:
\begin{equation}
u\,=\,(v-1)^{\theta +1/2}B(\theta +1, 1/2; w);\,\,\,\,\,w=\frac{
h_r-h_p}{h_0-h_p},
\label{h*-2}
\end{equation}
where $B(x, y; w)$ is incomplete $B$-function, defined by an integral:
\begin{equation}
B(x, y; w)\,=\,\int_0^w s^{x-1}(1-s)^{y-1}ds.
\label{incB}
\end{equation}
Thus, the ratio $w=\frac{h_r-h_p}{h_0-h_p}$ in this limit
is a function of only one variable $u/(v-1)^{\theta +1/2}$. For $h_0$
close to $h_{t2}$, the field $h_r$ is also close to $h_p$.
If $u/(v-1)^{\theta +1/2}\ll 1$, the ratio $w$ becomes small:
\begin{equation}
w\approx\left[\frac{\sqrt{2}u}{(\theta +1)(v-1)^{\theta+1/2}}\right]^
{{1\over\theta +1}}
\label{w}
\end{equation}
For completeness we present here equation for $h_r$ 
in the range $h_{t1}<h_0<h_{t2}$ without derivation:
\begin{equation}
u\,=\,(v-1)^{\theta +1/2}\left[2B(\theta +1, 1/2)\,-\,B(\theta +1, 1/2; w)
\right].
\label{h*3}
\end{equation}
If $u\ll 1$, but $v\gg 1$, the value $h_r$ depends on the
product $uv$. If it is small, then $h_r$ is still close to
$h_p$:
\begin{equation}
\frac{h_r-h_p}{h_p}\,=\,[(\theta +1)uv]^{1/(\theta +1)}\,=\,
\left[\frac{(\theta +1)\omega h_0L}{\gamma h_p^2}\right]^{1/(\theta +1)}.
\label{h*4}
\end{equation}
In the opposite limiting case  $(u\ll 1, uv\gg 1)$ the field $h_r$ is 
much larger than $h_p$, and the asymptotic $f(x)\sim x^{\theta}$ is 
not valid. In this case equation for $h_r$ reads:
\begin{equation}
h_r\,=\,\sqrt{\frac{\omega L}{\gamma}\left(2h_0-{\omega L\over\gamma}
\right)}.
\label{h*5}
\end{equation}
Still the reversal field $h_r$ is much less than the amplitude $h_0$.
For coercive field $h_c$ in the range $u\ll 1, v\gg 1$ one finds:
\begin{equation}
{h_c\over h_p}-1\,=\,\left({h_r\over h_p}-1\right)\cdot 2^{-1/(\theta
+1)};\,\,\,\,\,\,\,\,(uv\ll 1);
\label{h-c-1}
\end{equation}
\begin{equation}
{h_c\over h_p}\approx {1\over\sqrt{2}}.
\label{h-c-2}
\end{equation}
The shape and the area of the hysteresis loop depends on the
same parameters $u=\omega L/(\gamma h_p)$ and $v=h_0/h_p$. We consider
the range of small $u\ll 1$. If $v$ is close to one, the low-field
approximation for function $f(x)$ may be used in equation (\ref{z(h)})for $z$. 
Together with equation (\ref{magmoment}) it results in a
following relation:
\begin{equation}
{ M}\,=\,{M}_s\left[1\,-\,2\frac{
B\left(\theta +1, 1/2, 
\frac{h-h_p}{h_0-h_p}\right)}{B\left(\theta +1, 1/2,\frac{h_r-h_p}
{h_0-h_p}\right) }\right].
\label{magmoment1}
\end{equation}
For $-h_r<h<h_p$ on the lower branch of the hysteresis curve and
for $-h_p<h<h_r$ on the upper branch $ M=\pm M_s$. We presented
here only a comparatively simple case $h_0>h_{t2}$. The case $h_0<h_{t2}$ is
more complicated and we do not present here the explicit formulae for
magnetization in this range of amplitudes.

For  $u\ll 1$, $v\gg 1$ and $uv\ll 1$ we find in a similar way:
\begin{equation}
{M}\,=\,{M}_s\left[ 1\,-\,2\left(\frac{h-h_p}{h_r-h_p}
\right)^{\theta +1}\right]\,\,\,\,\,\,\,\,(h_p<|h|<h_r).
\label{magmoment2}
\end{equation}
and ${M}={M}_s$ for $h_r<|h|<h_0$.
Note that the hysteresis loop in this range of parameters is narrow,
i.e. $h_r, h_c\ll h_0$.

In the interval $u\ll 1, uv\gg 1$ the value of $h_r$, according
to eqn (\ref{h*5}), is much larger than
$h_p$. One can neglect the threshold field and employ the linear
asymptotic for $f(x)$ solving eqn.  (\ref{z(h)}). The result is:
\begin{equation}
Z(h)\,=\,{\gamma\over\omega}(h_0\,-\,\sqrt{h_0^2-h^2}).
\label{z1}
\end{equation}
and
\begin{equation}
{M}\,=\,{M}_s\left[1\,-\,2{\gamma\over\omega L}
(h_0\,-\,\sqrt{h_0^2-h^2})\right].
\label{magmoment3}
\end{equation}
The latter two equations are valid in the interval of fields 
$|h|<h_r$. Beyond of this interval the total magnetic
moment is fixed at its saturation value.
Note that a new characteristic length appears in the 
problem $L_{\omega}= \gamma h_0/\omega $.
Below we present results for the hysteresis loop area $\cal A$
in the range of small $u$ for different values of $v$ without
repeating of analogous calculations:\\
i) $v-1\ll 1$:
\begin{eqnarray}
{\cal A}&\approx&4{M}_sh_p\\
&\times&\left[ 1\,+\,\frac{\sqrt{2}(v-1)^{\theta +1/2}}
{u}\int_0^w B(\theta +1, 1/2; x)dx\right].
\label{A1}
\end{eqnarray}
ii) $v\gg 1, uv\ll 1$:
\begin{equation}
{\cal A}\,\approx\,4{ M}_sh_p\left[ 1+\frac{\left( (\theta +1)
uv\right)^{1/(\theta +1)}}{\theta +2}\right].
\label{A2}
\end{equation}
iii) $uv\gg 1$:\\
In this case $h_r\gg h_p$. Therefore all essential results 
formally coincide with those for $u\gg 1$ and the reader is referred to
a corresponding subsection.
Note that in the cases i) and ii) $\cal A$ is close to a constant
value ${\cal A}_0=4{ M}_sh_p$, but the deviation ${\cal A}-
{\cal A}_0$ scales with two parameters $u$ and $v$.

Now we proceed to a simpler case of large $u\gg 1$. In this case
the dynamic threshold field $h_{t1}$ is much larger than $h_p$:
\begin{equation}
h_{t1}\,=\,\frac{\omega L}{2\gamma}.
\label{h-t-3}
\end{equation}
Therefore, one can neglect $h_p$ when integrating eqn (\ref{z(h)}) and others.
Then we rediscover equation (\ref{z1}) for the coordinate. It implies
that
\begin{equation}
h_{t2}\,=\,\frac{\omega L}{\gamma}\,=\,2h_{t1}.
\label{h-1-1}
\end{equation}
Equation for $h_r$ coincides with (\ref{h*5}). Coercive field
$h_c$ is determined by equation:
\begin{equation}
h_c\,=\,\sqrt{h_{t1}(2h_0-h_{t1})}.
\label{h-c-3}
\end{equation}

Eqn. (\ref{magmoment3}) describes the shape of the hysteresis loop
for $h_0>h_{t2}$. It can be used also for the range of amplitudes
$h_{t1}<h_0<h_{t2}$ on the lower branch of the hysteresis curve $0<h<h_0$.
On the upper branch of this curve $h_r<h<h_0$ the sign of the
square root must be reversed.

The hysteresis loop area is determined by a following equation: 
\begin{eqnarray}
{\cal A}\,&\approx& 4{ M}_sh_r
 -{2{ M}_sh_r\gamma h_0\over \omega L}\\
&\times&\left(2 - \frac{\arcsin{{h_r\over h_0}}}
{\sqrt{1-(h_r/h_0)^2}}
- \sqrt{1-({h_r\over h_0})^2}\right).
\label{A3}
\end{eqnarray}
This equation is valid for $h_0>h_{t2}$.
If, on the contrary, $h_0<h_{t2}$, the area is:
\begin{eqnarray}
{\cal A}&=&2{ M}_s
\int_0^{h_r}\left( 1-{Z(h)\over L}
\right)dh\\
&+&2{ M}_s\int_{h_r}^{h_0}\frac{Z_+(h)-Z_-(h)}{L}dh
\label{A-sofist}
\end{eqnarray}
where $Z_{\pm}(h)$ denote the value of coordinate on the upper ($+$)
and lower ($-$) branches of the hysteresis curve respectively.
Performing the integration, we find:
\begin{equation}
{\cal A}\,=\,4{M}_sh_r\left(1-{\gamma h_0\over\omega L}
\right)\,+\,\pi{M}_sh_0{\gamma h_0\over\omega L}.
\label{A-sofist1}
\end{equation}
One can observe that for the case $u\gg 1$ the static 
threshold field $h_p$ does
not enter physical results. Instead a new dimensionless parameter
\begin{equation}
u^{\prime}=\omega L/(\gamma h_0) = L/L_{\omega}
\label{uprime}
\end{equation}
emerges. In the range of existence
of the full hysteresis loop $u^{\prime}<1/2$. 
Therefore $u^{\prime}$ can not be
large. However, it can be very small. It is worthwhile to consider
separately the behavior of all values of interest in this asymptotic
regime (large fields or small frequencies). For $u^{\prime}\ll 1$
we find:
\begin{equation}
h_r\,\approx\,h_c\,=\,h_0\sqrt{u^{\prime}};
\label{h-*-c}
\end{equation}
\begin{equation}
{M}(h)\,=\,{M}_s\left( 1\,-\,{2\over u^{\prime}}\frac{h^2}{h_0^2}
\right);
\label{magmoment4}
\end{equation}
\begin{equation}
{\cal A}\,\approx\,4{M}_sh_r.
\label{A4}
\end{equation}
Near the threshold field $h_{t1}$ the values of $h_r$
and $\cal A$ are given by following asymptotics:
\begin{eqnarray}
h_r&\approx&2\sqrt{h_{t1}(2h_0-h_{t1})}; \\
{\cal A}&=& {\pi{ M}_s h_{t1}\over 2} 
+ 4\pi{ M}_s\sqrt{h_{t1}(2h_0-h_{t1})}.
\label{asimp}
\end{eqnarray}
Near the second threshold field $h_{t2}$
the reversal field $h_r$ is close to $h_0$:
\begin{equation}
h_r\,\approx\,h_0\, - {(h_0 - h_{t2})^2 \over 2 h_{t2}}.
\label{appr2}
\end{equation}
For $u^{\prime}\ll 1$ we find scaling behavior of all hysteresis
characteristics with universal critical exponents, independent on
the pinning centers. Therefore, one can expect that the same scaling
is valid for a clean ferromagnet with the relaxational dynamics.
In particular, we have found that 
\begin{equation}
{\cal A}\propto\omega^{1/2}h_0^{1/2}
\label{ares}
\end{equation}
In this section
we assumed  that the magnetization in a single domain reaches
its saturation value. If it is not the case, the magnetization on 
the parts of hysteresis curve beyond the hysteresis loop (``the whiskers'')
is not a constant. The exact shape of the hysteresis curve in this
case depends on the single-domain equilibrium magnetization
curve.

\section{Creation of bubbles by a moving domain wall}
\label{traps} 
In this section we show that, at sufficiently high magnetic field 
$h_p < h_0 <\eta$, the moving wall passes a series of defects which it 
cannot overcome. Instead it leaves closed domains of magnetization, opposite 
to the propagating magnetization, which serve as nuclei at the 
next half-cycle  of the hysteresis process.

So far we assumed that an almost planar domain wall is pinned by 
typical fluctuations of the random field $\eta (\vec r)$. In a (spherical)
region of the volume $R^d, d=D+1$ a typical fluctuation of 
$\int_{R^d}\eta (\vec r){\rm d}^dr$ is of the order $\pm\eta (l^dR^d)^{1/2}$,
the corresponding probability is of the order 1/2. 
In the following we will consider the possibility of clusters of rare 
random field fluctuations which may serve as nuclei of reversed spins when 
the field direction changes. 
A cluster of the size $R$ consisting mainly 
minimal negative value $-\eta$ of the 
random field  has the probability 
$p(1,R)\simeq \exp{(-(R/l)^D\ln{2})}$. Such a cluster
pins the domain wall  for  $h<\eta$ even if  $h >h_p$.
In general, if the concentration of 
$-\eta$ sites in a cluster is $c$, its probability $p(c,R)$ is of the order
$\exp{\left[ -(R/l)^d[c\ln{c} +(1-c)\ln{(1-c)}+\ln{2}]\right]}$ 
and the pinning force density is of the order $(2c-1)\eta$.
Since $h_p\ll \eta$, there is a field region $h_p<h<(2c-1)\eta$
in which the domain wall can be pinned locally by these rare fluctuations of 
the random field. The true pinning condition is now given by 
(see also \cite{EbEngel} for similar considerations)
\begin{equation}
\frac{\Gamma}{R}+h-(2c-1)\eta <0
\label{eq:13}
\end{equation}
where the first term denotes the Laplacian curvature 
force. Thus, the domain wall 
cannot overcome the pinning cluster if it has the size $R$ bigger than 
$R_{min}$, where 
\begin{equation}
R_{\rm min}\approx\frac{\Gamma}{(2c-1)\eta -h}\;.
\label{eq:Lmin}
\end{equation}
Using (\ref{eq:Lmin}) we get
\begin{equation}
p(c,R_{\rm min})\approx 
\exp{\left\{-\left[ 
\frac{\Gamma}{(2c-1)\eta l -hl}\right]^d
g(c)
\right\}}
\label{eq:p-c-Lmin}
\end{equation}
where
\begin{equation}
g(c) =  c\ln{c} +(1-c)\ln{(1-c)}+\ln{2}
\label{eq:p-c-Lmin1}
\end{equation}
A closer inspection shows, that in $d\ge 2$ dimensions $p(c, R_{\rm min})$
has its maximum at $c=1$. The mean distance between these strong 
pinning clusters is therefore of the order
\begin{equation}
L_{cluster}\approx l\exp{\left\{\left[\frac{\Gamma}{(\eta -h_0)l}\right]^d
\frac{\ln{2}}{d}\right\}}
\label{eq:R-c}
\end{equation}
Here we have replaced $h$ by by the maximum field strength during one cycle 
$h_0$. 
Since the interface cannot overcome the pinning cluster it will surround
it and finally leave
 it behind as an island encircled by a domain
wall (Fig. \ref{nat2}). These islands serve 
as nuclei of the favorite phase once 
the external field is reversed. 
Thus, in the expressions found in Sec.\ref{line}
the system size $L$ has to be replaced by 
$L_{cluster}$ as soon as $L>L_{cluster}$.
If $h_0$ approaches and finally exceeds $\eta $, $L_{cluster}$ diverges 
and hence this type of 
magnetization reversal process disappears. Domain walls originate then 
either from surfaces or from a nucleation process which we consider in the 
next section.

\section{ Nucleation controlled hysteresis}\label{nucleation}
This mechanism works when the nucleation is the longest process. Let there
exists a number of defects or impurities which lower the energy barrier for
reversing magnetization locally. Still the reduced barrier $\Delta$
remains rather large so that the average reversal time 
$\tau_r(0)=\nu^{-1}\exp{(\Delta /T)}$ ($\nu$ is a microscopic frequency)
is very large in comparison to the oscillation period:
$\omega \tau_r\gg 1$. However, the reversal time depends on magnetic field:
\begin{equation}
\tau_r(h)=\tau_r(0)\exp{\left( -\frac{{ M}_ah}{T}\right)}
=\nu^{-1}\exp{\left(\frac{\Delta -{ M}_ah}{T}\right)}
\label{eq:t-H}
\end{equation}
where ${M}_a$ is the saturation magnetic moment in the activation volume.
We assume also that $\omega /\nu\ll 1$, but according to the accepted
assumption $\exp{(\Delta /T)}\gg \nu /\omega$. Then the probability of spin
reversal in an individual nucleus is negligibly small if 
 $\frac{{M}_ah_0}{T}<\frac{\Delta}{T}-
\ln{\biggl(2\pi\frac{\nu}{\omega}\biggr)}$.
It becomes reasonably large starting from the dynamical threshold for 
the amplitude $h_0$:
\begin{equation}
h_{t1}=\frac{\Delta}{{M}_a}
-\frac{T}{{M}_a}\ln{\biggl(2\pi\frac{\nu}{\omega}\biggr)}
\label{eq:H-t-a}
\end{equation}
Due to the sharpness of the exponential function, the reversal time
$\tau_r(h)$ fastly becomes much smaller than the oscillation period
 $2\pi /\omega$ as soon as $h(t)$ overcomes the threshold values $h_{t1}$.
With the precision of a small parameter $\nu /\omega\exp{(-\Delta /T)}$ 
the reversal proceeds at a fixed field $h=h_{t1}$ which simultaneously 
plays the role of
coercive field $h_c$ and the reversal field $h_r$ (see section
\ref{line}). With the same accuracy the reversal proceeds at a moment of time 
 $t_r=\omega^{-1}\arcsin{(h_{t1}/h_0)}$. The hysteresis loop has a
rectangular shape (see Fig.\ref{fig:loop}). With a little 
higher accuracy the transition 
proceeds at a value of the phase $\varphi =\omega t$ determined by
\begin{equation}
\frac{{M}_ah_c}{T}\sin{\varphi}=\frac{\Delta}{T}-\ln{\left(
\frac{2\pi\nu}{\omega}\varphi\right)}
\label{eq:coreq}
\end{equation}
which results in a corrected value of the coercive field:
\begin{equation}
h_c=h_r=h_{t1}-\frac{T}{{M}_a}\ln{\left(\arcsin{\frac{h_{t1}}{h_0}}\right)}
\label{eq:Ha-c-a}
\end{equation}
As soon as the magnetization reverses 
in an individual nucleus, its walls expand 
fastly and, for a much shorter time than $t_r$, the full magnetic
moment reverses. The case in which the propagation of domain walls 
is the longest process has been considered earlier in Sections \ref{line}
and \ref{traps}. The HLA for a rectangular loop is simply:
\begin{eqnarray}
{\cal A}&=&4{M}_sh_c\\
&\approx& 4\frac{{M}_s}{{M}_a}
\Delta\left[ 1-\frac{T}{\Delta}\ln{\frac{2\pi\nu}{\omega}}-
\frac{T}{\Delta}\ln{\left(\arcsin{\frac{h_{t1}}{h_0}}\right)}\right]
\label{eq:A-a}
\end{eqnarray}
Since  $M_a h_{t1}\ll \Delta$ the value 
$1 - (T/\Delta)\log {2\pi\nu /\omega} \ll 1$.
For $h_0\gg h_{t1}$ the HLA reads:
\begin{equation}
{\cal A}={\cal A}_0\left[ 1-\frac{T}{\Delta}\ln{\frac{2\pi\nu}{\omega}}
+\frac{T}{\Delta}\ln{\frac{h_0}{h_{t1}}}\right]
\label{eq:A-a-1}
\end{equation}
Note that at fixed $\omega$ and varying $h_0$ 
we find 
\begin{equation}
{\cal A}={\cal A}_0(\omega )
\left[ 1-\frac{T}{M_a h_{t1}}\ln{\frac{h_0}{h_{t1}}}\right]
\label{eq:A-a-2}
\end{equation}
and  $T/(M_a h_{t1})\gg T/\Delta$.
This theoretical result can be compared with recent measurements by
Suen and Erskine \cite{erskine}. They have measured the HLA for a 
thin film of $Fe/W(110)$ in a wide range of frequencies
 $\omega$ and amplitudes $h_0$ . They concluded that 
 ${\cal A}\sim h_0^{\alpha}\omega^{\beta}$ with small exponents 
 $\alpha \sim 0.2$ and very small exponent 
 $\beta\sim 0.03-0.09$. Both exponents depend on temperature. 
These conclusions 
are compatible with our result (\ref{eq:A-a-1}). Indeed, at such small 
values of exponents it is impossible to distinguish between the 
logarithmic dependence and the exponential dependence with a small 
exponent. Our predictions $\alpha =T/{M}_ah_{t1}$ and $\beta =T/\Delta$
agree with the experimentally discovered linear temperature dependence of
the exponent $\beta$. It would be worthwhile to measure the frequency
dependence of the dynamical threshold 
field $h_{t1}$. The fact of the existence 
of such a threshold has been observed in the same experiment. It should 
be noted that the value $T/{M}_ah_{t1}$ is not so small in the
experiment $\sim 1/4$. Therefore we expect that our theory is correct
with the accuracy 25\% only. On the other hand, the statement that
there exists the power-like dependence of the HLA on $h_0$ is not
proved convincingly by this experiment since the interval of
variation for $h_0$ was too small (the amplitude changed about 10 times).

So far we considered $\Delta$ and ${M}_a$ as fixed values. This is 
correct if only one type of defects mediates the nucleation. More
realistically these values are random. Then the defects with the minimal 
ratio $\Delta /{M}_a$ initiate the magnetization reversal process.

According to Eqn.(\ref{eq:H-t-a}) $h_{t1}$ becomes zero at very 
small frequencies $\omega\le 2\pi\nu\exp{(-\Delta /T)}$. Our theory
is not valid for such a small frequency, but it reflects correctly the 
narrowing of the HL. On the other hand, the transition is smeared
over the interval of magnetic field $\Delta h \sim T/{M}_a$. It is
small in comparison to $h$ if $h_0{M}_a\gg T$. At a smaller
amplitude $h_0$ the reversal is possible, but fluctuations of
magnetization grow rapidly and become of the order of magnetization itself.
In principle, it is possible that the nucleation time is of the same 
order of magnitude that the time of domain wall propagation. However, since
the nucleation time depends fastly on parameters, a small change of the 
regime makes one of the two times much larger than another. Thus, 
presumably the nucleation mediated hysteresis loop has the rectangular
shape, whereas a curved shape indicates that the hysteresis is associated
with the domain wall propagation or with strong fluctuations of magnetization.

A reason for the rounding of the HL in the case of the nucleation
controlled hysteresis can be that the sample is splitted into a number
of magnetically disconnected grains with different $\Delta$ and
${M}_a$ in each grain. Then the shape of the hysteresis loop
reflects the distribution function for the ratio $\Delta/{M}_a$.

\section{The Adiabatic Dynamics}
\label{adiabatic}
In the case of strong Ising anisotropy the dynamic
equation for magnetization $m$ in the  continuum limit 
has the form:
\begin{equation}
\label{HD} 
{\partial m\over \partial t}= - \Gamma {\partial F\over \partial  m}
\end{equation} 
where $F$ is a free energy. In the adiabatic limit the total
magnetization $m(t)$ at the moment $t$ can be represented as
\begin{equation}
\label{m} 
m(t) = m_0(T,h(t)) + m_1(t)
\end{equation} 
where $ m_0(T,h(t)) $ is the equilibrium value of magnetization for 
a given 
momentarily value of the 
magnetic field $h(t)$ and $m_1(t)$ is the deviation 
of magnetization from its equilibrium value.
We discuss later  restrictions imposed by
the adiabaticity condition. Substituting $m(t)$ into Eq.(\ref{HD}),
one finds equation for $m_1$:
\begin{equation}
\label{hi} 
{\chi \dot h}= - \Gamma {\partial^2 F\over\partial m^2} m_1,
\end{equation} 
where $\chi ={\partial m/\partial h}$ 
is the magnetic susceptibility. Eq.\ref{hi} can be
rewritten in a form:
\begin{equation}
\label{m1} 
m_1  = - {\chi^2 \over \Gamma} \dot h
\end{equation}
The hysteresis appears since the derivative $\dot{h}$ changes its
sign when $h$ passes its extremal values $\pm h_0$. 
For harmonically-oscillating field $h(t) = h_0 \sin\omega t$,
its derivative may be expressed in terms of the field itself:
\begin{equation}
\label{h1} 
\dot h  = \pm \omega \sqrt{h^2_0 - h_c^2}.
\end{equation} 
Plugging (\ref{h1}) into (\ref{m1}), we find:
\begin{equation}
\label{m2} 
m_1  = \mp {\omega \chi^2 \over \Gamma}  \sqrt{h^2_0 - h^2}
\end{equation} 
Perturbation theory is valid if $m_1 \ll (m_0)_{max} = m_0(T,h_0)$,
i.e. if $\omega\chi (h_0) /\Gamma \ll 1$. The coercive magnetic field $h_c$ 
satisfies a following equation:
\begin{equation}
\label{ma} 
m_0(T,h_c)=\vert m_1(h_0,h_c,T)\vert ={\omega \chi^2 \over \Gamma}\sqrt{h^2_0 - h^2_c}
\end{equation} 
Assuming $h_c \ll h_0$, eqn. (\ref{ma}) can be transformed as follows:
\begin{equation}
\label{m01} 
m_0(T,h_c) = {\omega \chi^2 h_0\over \Gamma}
\end{equation}
The HLA ${\cal A}$ can be calculated by knowledge
of $m_1$:
\begin{equation}
\label{area2}
{\cal A}=4\int_0^{h_0}m_1dh=4\omega\int_0^{h_0}{\chi^2\over\Gamma}
\sqrt{h_0^2-h^2}dh.
\end{equation}
Near Curie point the relaxation time $\tau_c\propto\Gamma^{-1}\chi$
grows as $\tau_c\sim\tau_0\epsilon^{-\nu z}$ where
$\epsilon =(T-T_c)/T_c$, $\nu$ is the correlation length
critical exponent ($\nu =1$ for 2D Ising model) and $z$ is
the dynamical critical exponent. The best numerical value for
$z$ is $z=2.17$ (for a modern review see
\cite{adler} and references therein). If $\epsilon =0$ and
$h\neq 0$, the relaxation time is proportional to $h^{8z/15}$.
For adiabaticity it is necessary that $\omega\ll\omega_c=
max(\tau_0^{-1}\epsilon^z, \tau_0^{-1}(h_0/h_{ex})^{8z/15})$, where
$h_{ex}$ is the saturation (exchange) field. 

The amplitude $h_0$
should be considered as small (the case i) or large (the case ii)
depending on its ratio to a characteristic field $h^{\ast}(\epsilon)=
h_{ex}\epsilon^{15/8}$. For a small amplitude $h_0\ll h^{\ast}(\epsilon)$
one can employ linear approximation for $m_0(T,h)=\chi (\epsilon)h$
to solve eqn.(\ref{m01}). Then $h_c=h_0\omega\chi(\epsilon)/
\Gamma(\epsilon)$. Next we use the relation: 
$\chi (\epsilon)/\Gamma \sim \tau_0\epsilon^{-z}$ 
to find:
\begin{equation}
\label{h-c-scale}
h_c\,=\,\omega h_0\tau_0\epsilon^{-z}=\,\omega h_0\tau_0\epsilon^{-2.17}
\end{equation}
For a large amplitude $h_0\gg h^{\ast}(\epsilon )$ a dimensionless ratio
$\kappa =\omega\chi (\epsilon)/\Gamma (\epsilon )$ matters. If it is
much less than the large value $h_0/h^{\ast}$, then the coercive force
can be calculated as $h_c=\kappa h_0$, exactly as in eqn. 
(\ref{h-c-scale}). Otherwise $h_c\gg h^{\ast}$. Then an approximate solution
of eqn. (\ref{m01}) reads:
\begin{equation}
\label{h-c-high}
h_c\approx h_{ex}\left(\omega\tau_0
\frac{h_0}{h_{ex}}\right)^{-\frac{15}{15+8z}}
\sim h_{ex}\left(\omega\tau_0\frac{h_0}{h_{ex}}\right)^{-0.46}
\end{equation}
Corresponding results for the HLA are
as follows:
\begin{equation}
i)\,\,\,\,\,\,\,\,{\cal A}\propto\omega h_0^2\frac{\chi^2(\epsilon )}
{\Gamma (\epsilon )}\sim{\omega h_0^2}{\tau_0}\epsilon^{-z-\gamma}
\sim{\omega h_0^2}{\tau_0}\epsilon^{-3.92}.
\label{A21}
\end{equation}
\begin{equation}
\label{A22}
ii)\,\,\,\,\,\,\,\,{\cal A}\propto\omega h_0\frac{h^{\ast}(\epsilon )
\chi^2(\epsilon )}{\Gamma
(\epsilon )}={\omega h_0h_{ex}}{\tau_0}\epsilon^{-2.04}.
\end{equation}

Though we can not calculate explicitly the hysteresis loop
area in the opposite, anti-adiabatic regime $\omega_c\ll\omega\ll J/\hbar$,
the dimensionality arguments lead to a necessary estimate. Indeed,
according to general equation (\ref{area2}), its scaling dimensionality
is 
\begin{equation}
{\cal A}\propto\chi h_0^2,
\end{equation}
where we have used the facts that $\omega$ and $\Gamma/\chi$ have the same
dimensionality and that 
$h_0\ll h^{\ast}(\omega )=h_{ex}(\hbar\omega/J)^{8z/15}$.
Thus,
\begin{equation}
\label{anti-ad}
{\cal A}\propto\chi(\omega )h_0^2\propto\omega^{-\gamma /z}h_0^2=
\omega^{-0.8}h_0^2.
\end{equation}
Typical orders of magnitude are: $\tau_0^{-1}\sim J/\hbar\sim 4\cdot 
10^{13} Hz; \epsilon\sim 10^{-3}; \omega_c(\epsilon )\sim 2 MHz;
h^{\ast}(\epsilon )\sim 3 Oe; T_c\sim 300K; T-T_c\sim 0.3K.$ 

Note that all calculations in this section have been made for $T>T_c$,
so that we could ignore the problem of nucleation and domain wall
motion. Below the Curie temperature the problem of spinodal decomposition
appears, when the field reaches the spinodal. Then, again, either the
nucleation or the domain wall motion prevails in the magnetization
reversal. It does not happen if either $h_0\gg h^{\ast}(\epsilon )$
or $\omega\gg\tau^{-1}_c$. Then one can use previous results. However, if
these conditions are not met, then either nucleation or the domain
wall propagation prevails in the magnetization reversal process.
To establish, what is more important, let us consider corresponding
time scales. The nucleation time for a pure sample is:
\begin{equation}
\tau_r\,=\,\tau_c(\epsilon)\cdot
\exp{\left({h^{\ast}(\epsilon)\over h}\right)}.
\label{t-r-c} 
\end{equation}
Though this time scale grows at approaching to the Curie temperature,
it can be sufficiently small. Its minimization over $\epsilon$ at 
fixed value of $h$ gives:
\begin{equation}
\tau_r\propto\tau_0\left({h_{ex}\over h}\right)^{z/(\beta +\gamma )}
\,=\,\tau_0\left({h_{ex}\over h}\right)^{1.157}
\label{t-r-min}
\end{equation}
For the exchange field about $10^6Oe$ and external field about $100Oe$
the nucleation time becomes of the order of $10^{-9}sec$. Thus, the
nucleation proceeds very fastly and the process of domain wall 
propagation prevails in this range of fields and frequencies. This
conclusion explains the results of our numerical calculations for
temperatures close to the Curie point (see section \ref{mc}).

\section{Simulation of the Domain Growth}
\label{mc}
In this Section the process of magnetization reversal
in a few monolayers thick magnetic film
is modeled by analogous process in the two-dimensional
Ising model with the Glauber dynamics. 
In simulation we neglect
the dipole-dipole interaction and 
demagnetization effects.
The Hamiltonian of the Ising system in a 
time-dependent field  $h(t)$  has a form:
\begin{equation}
\label{H} 
 {\cal H}_b = -{1\over 2}  \sum_{\bf r, a}
 J_{\bf a}({\bf r}) \sigma_{\bf r}\sigma_{\bf r+a}
 -  h(t) \sum_{\bf r} \sigma_{\bf r}, 
\end{equation} 
where summation over ${\bf r}$ runs 
over the lattice sites
and ${\bf a}$ labels nearest 
neighbors. In a perfect
ferromagnetic film $J_{\bf a}({\bf r}) = J_0> 0$
independently on $\bf r$.
In the case of quenched disorder
we assume that the exchange integral is ferromagnetic 
($J_{\bf a}({\bf r}) = J_0 > 0$) with the probability $1-\vartheta$ 
and antiferromagnetic ($J_{\bf a}({\bf r}) =- J_0 < 0$)
with the probability $\vartheta$ ( $\vartheta \ll 1$).
No correlation in the location of random bonds is assumed. 
In Sec.\ref{dw} it is shown that this  type of disorder 
is of general importance for the problem of 
the domain wall moving in the disordered media.
Weak random bond disorder 
does not destroy the long-range order in the 2D Ising model.  
The magnetic field $h(t)$  is supposed to oscillate harmonically as 
 $h(t) = h_0\sin{(\omega t)}$, unless a different assumption 
is not specially formulated.
The process of magnetization reversal can be
divided into two stages: 
nucleation of domains with opposite magnetization 
and growth of these domains. 
Which process dominates depends on 
system parameters and its history. 
In this Section we consider a limit
of the nucleation time much smaller 
than the growth time. Then 
the magnetization reversal  process is
dominated by the domain wall propagation
 discussed in Sec.\ref{dw} and \ref{line}. 

We employed  the
Monte-Carlo simulation with the Glauber 
dynamics (see e.g. \cite{bin})
to check Eq.\ref{eq:6} for perfect systems 
and systems with disorder. 
We modelled the disorder by 
a small concentration
of randomly distributed  antiferromagnetic bonds. 
This disorder weakly influences the phase
diagram and results only in a small shift of $T_c$,
where $T_c$ is an Ising transition
temperature of the model Eq.\ref{H} with $h=0$. 
We have found the linear dependence
of domain wall velocity 
on applied magnetic field at all relevant temperatures
with and without disorder. 
Apparently, in these simulations we 
dealt with the limit
of weak pinning, so that $h_p \ll h_0$.
However, at low temperatures ($T < 0.1\cdot T_c$)
even a  small disorder, which almost does not change the  
thermodynamics, modifies the  domain wall dynamics  drastically.
The results related to  the domain wall motion
and hysteresis phenomena 
in a weakly disordered system at low temperature
will be discussed elsewhere.

The domain growth was studied  in two 
geometries: stripes and  circles. 
To distinguish the  domain growth from domain 
nucleation we examined either 
a specially prepared defect for
fast nucleation, or a prepared nucleus
with the opposite direction of  
magnetization possessing the shape of the circle or stripe.
In both cases the magnetization changed  in accordance 
to the model of a straight DW, i.e. similar 
to Eq.\ref{magmoment3}.

Simulated hysteresis loops for different values of 
magnetic field amplitudes ($h_0 = 0.025 J$,  $h_0 = 0.05 J$, 
and $h_0 = 0.125 J$)   
and constant period of $ 8\cdot 10^{3} $ Monte-Carlo steps 
are shown in Fig. \ref{f2}.
Hysteresis loops for different periodes ( $ 4\cdot 10^{3} $,
$ 8\cdot 10^{3} $ and  $ 256\cdot 10^{3} $  Monte-Carlo steps) 
and constant  $h_0=0.05J$ are shown in Fig.\ref{f2a}.  
The observed hysteresis loop behavior
obey the general classification scheme developed
in Secs. \ref{dw},\ref{line} (compare with Fig. \ref{fig:loop}).  
In particular, the asymmetric   loops in  Figs. \ref{f2},\ref{f2a}
correspond to the loop  in  Fig.\ref{fig:loop}a,
the loops with whiskers  in  Figs. \ref{f2},\ref{f2a}
corresponds to the loop  in  Fig.\ref{fig:loop}d,
and the symmetric loops without whiskers  in  Figs. \ref{f2},\ref{f2a}
corresponds to the loop  in  Fig.\ref{fig:loop}c.
The concentration of random bonds 
in the system was $\approx 0.8\%$.

Fig.\ref{f6} show the time sequence of inflating
domains  for the system with $T=0.5T_c$ and $h_0 = 0.25 J$.
The concentration of random bonds 
in this case was $\approx 3\%$.
The nucleation first proceeds at three regions, 
then the corresponding
domains inflate rapidly and two of them merge. 
The time development of the inflation
process is shown in Fig.\ref{f6}.
The total magnetization  for each snapshot 
is indicated in the Fig. \ref{f6}.
The analysis show that the 
inflation of a domain with reversed
magnetization is described by simple 
law Eq.\ref{magmoment3}.

Let $L_N$ be
the mean distance between nucleation centers.
It plays the same role 
as the system size  $L$ in  Sec.\ref{line}.
The ratio $L_N/L_{\omega} = u_{N}$ 
is an analog of the  dimensionless  
parameter $u^{\prime}$ introduced by Eq.\ref{uprime}.
In the limit $L_N \ll L_{\omega}$ (compare Sec.\ref{line})
the domain growth will be fast enough to provide the full 
hysteresis cycle with $h_r\ll h_0$ (see  Fig.\ref{fig:loop}d ).
We evaluated the 
scaling behavior of  $h_c$ and $\cal A$
as a function of  $\omega$ and $h_0$ 
in the limit  $L_N \ll L_{\omega}$ using  Eq.\ref{z1}.
To deal with the domain wall propagation dominated hysteresis,
we considered temperature close to Curie point.
We argued earlier (see Sec.\ref{adiabatic}) that in this case the 
nucleation proceeds fastly enough.
Fig.\ref{f1} shows domain distribution
in the applied reversed magnetic field at $T=0.95 T_c$.
Multiple nuclei which are seen on this picture 
serve as evidence that the nucleation is really fast enough.

At  $h=h_c$ the typical domain size $L_d$ is
of the order of the mean distance between domains.
Since $L_N \ll L_{\omega}$ we conclude 
that $h_c \ll h_0$, and  $L_N \approx  (\gamma /h_0 \omega)(h_c)^2$.
Assuming  that the average distance between 
nucleation centers does not depend significantly on $\omega$,
we immediately obtain that the coercive field 
$h_{c}$ scales as  
$h_{c} \propto \sqrt{\omega}$ (compare with   Eq.\ref{h-*-c}).
If, in addition,  the average distance between 
nucleation centers does not depend significantly on $h_0$,
we have  $h_{c} \propto \sqrt{h_0\omega}$.
If we assume that the 
magnetic field is strong
enough and saturation magnetization does not
change significantly by changing $h_0$,
the HLA area scales in the
same way as  $h_{c} $, i.e. ${\cal A} \propto \sqrt{h_0\omega}$. 
This is a simplified qualitative version of arguments leading to
Eq.\ref{ares} of Sec.\ref{line}

Fig.\ref{f7} shows the dependence of the hysteresis loop area
${\cal A}$  on $\omega $ in the double 
logarithmic coordinates at $T=T_c$.  The $\sqrt{\omega} $
scaling behavior is valid over three decades
of the $\omega $ variation. However the
the range of magnetic field values available
is only one decade. For this reason we have
only checked that magnetic field dependence 
is consistent with  
${\cal A} \propto \sqrt{h_0} $ dependence.

To check the proposed mechanism of magnetization
reversal we have also
simulated this process for the 
magnetic field  $h= h_0\sin^3(\omega t)$.
In this case both $h_c$ and $\cal A$ 
scales in accordance with the analytical results as  $\omega^{3/4} $.

Though we put $T=T_c$, the system is far away from the
critical point due to relatively large magnetic field. 
Indeed, one can estimate $\epsilon $ defined in 
Sec.\ref{adiabatic} from the relation  
$\epsilon \approx ({h_0/h_{ex}})^{8/15}$  
(see Sec.\ref{adiabatic}). In our simulations
$h_0/h_{ex}= 0.05-0.005$ which give   $\epsilon \approx 0.2 - 0.06$.
It means that the deviation from the 
critical point is large enough.
With these values of fields our system is also far away
from the strong fluctuation regime
described in Sec.\ref{adiabatic}. 
A necessary condition for this regime is that $m_0 \ll 1$.
The magnetic  moment $m_0$ scales 
as $m_0 \propto ({h/h_{ex}})^{1/15}$. 
Therefore the field  $h_0 \approx 0.005h_{ex}$ is rather strong
it gives $m_0 \approx 0.7$.

With the temperature decreasing, away from critical
region the HLA deviates strongly from the power-law behavior
as it is follows from Fig.\ref{f7}.
We expect that
the HL  changes its shape and characteristics from 
those dominated by domain wall propagation 
to those  dominated by nucleation.
The crossover behavior can not be 
described by a single power law.

\section{Conclusions.}\label{conclusions}
We studied main mechanisms of the hysteresis in a ferromagnet:
the driven DW motion, nucleation and retardation in non-linear
magnetization dynamics. Each of them may be dominant at proper
conditions. A process is dominant if it provides maximum 
value for the coercive field $h_c$.

On the background of general theory of the interface motion in
a random medium we studied the hysteresis process controlled
by the DW motion. We introduced two dynamic threshold fields
$h_{t1}$ and $h_{t2}$ corresponding to the occurrence of the full
magnetization reversal and to the occurrence of the single-valued
parts on the hysteresis curve respectively. These dynamical threshold fields 
are larger than the static threshold field $h_p$ which is required
to start a motion of the DW. We established that $h_{t1}$ and
$h_{t2}$ measured  in units $h_p$ are functions of one dimensionless
parameter $u=\omega L/\gamma h_p$ (see notations in sect. 
\ref{introduction}, \ref{dw}, \ref{line}). The coercive
field $h_c$ and the reversal field $h_r$, measured in the
same units, can be expressed in terms of one function of two
dimensionless parameters:
\begin{equation}
h_c\,=\,h_pF\left(\frac{\omega L}{2\gamma h_p}, \frac{h_0}{h_p}
\right);\,\,\,\,\,
h_r\,=\,h_pF\left(\frac{\omega L}{\gamma h_p}, \frac{h_0}{h_p}
\right).
\label{h-scale}
\end{equation} 
Experimental observation of this type of scaling would be the best
indirect evidence of the DW motion controlled hysteresis.
The direct observation of the DW motion by SMOKE and STM methods
in principle is possible
\cite{nikitenko} and is much appreciate. At large fields 
$h_0\gg h_p$ the defects are inessential. Therefore, the dependence
on $h_p$ must vanish from scaling laws (\ref{h-scale}). It happens
indeed, and both fields $h_c$ and $h_r$ are expressed in terms of
one dimensionless parameter:
\begin{equation}
h_c\,=\,h_0F\left(\frac{\omega L}{2\gamma h_0}\right);\,\,\,\,\,\,
h_r\,=\,h_0F\left(\frac{\omega L}{\gamma h_0}\right),
\label{scale-large}
\end{equation} 
where $F(x)=2\sqrt{x(1-x)}$. Note that in this limit 
the value $h_p$ vanishes
from equations for the dynamic threshold fields as well:
$h_{t1}=\omega L/\gamma$, $h_{t2}=2h_{t1}$. We presented also corresponding
equations for the HLA to which the most experimental efforts were
concentrated. However, we would like to emphasize that the HLA is not
the only measurable characteristics of the HL and even not the
most informative its characteristics: the fields $h_{t1}, h_{t2}, h_c,
h_r$ as well as the shape of the hysteresis curve are not less
interesting. The functional dependence of the fields $h_{t1}$ and 
$h_{t2}$ on $h_0$ at $h_0$ not much overcoming the static threshold
field $h_p$ allows to restore the basic function $f((h-h_p)/h_p)$
of theory of the DW motion in a defect medium.

For sufficiently large $h_0$ the length $L$ in equations
(\ref{h-scale}, \ref{scale-large}) and others is the size of
the system. Thus, the size effect is observable in the HL
characteristics. However, at fields, larger than $h_p$, but
still smaller than the maximum defect strength, the moving DW
produces bubbles playing the role of ready nuclei
for magnetization reversal at the next half-cycle of the 
hysteresis loop. Thus, the length $L$ in this case is the
average distance between the bubbles. Its strong dependence
on the the field amplitude $h_0$ makes the scaling laws
(\ref{h-scale}, \ref{scale-large}) less transparent, but
they become size-independent. The scaling becomes simple
again if the dominant defects are of topographic origin.
The  density  of topographic defects is 
independent on the amplitude of magnetic field. Then $L$ is the
average distance between such defects.

In the case, when the driven DW are almost free ($h_0\gg h_p$)
and the HL is narrow ($h_0\gg h_{t1}$), the HLA was found to be
proportional to $\omega^{1/2}h_0^{1/2}$. This conclusion is
supported by our numerical MC simulation.

In the nucleation controlled process 
almost rectangular HL shape is expected,
unless the sample is divided into a multitude
of magnetically disconnected grains. Note that the shape of
almost vertical parts of the HL in this 
case is determined by the DW motion
and, therefore, is universal. In the case of nucleation controlled
hysteresis the HLA must grow logarithmically with the frequency
and  amplitude of  magnetic field. 
However, the coefficient at the
logarithm of frequency is smaller than that at the logarithm of
the amplitude. 
Slow nonlinear critical dynamics may produce the HL with various
scaling limits as described in sect.\ref{adiabatic}. 

The comparison with the experiment is still rather poor since
there is no systematic study of the HL characteristics as 
functions of dimensionless parameters. The experimental efforts
were focused on the HLA, but the precision of the measurements 
is not high enough for reliable determination of scaling.
Sue and Erskine \cite{jim}
reported the observation of linear dependence of the HLA on the
logarithm of frequency \cite{footnote} with the coefficient proportional
to temperature. Both these facts agree with our theory. However,
they also claimed the power dependence of the HLA on the amplitude
$h_0$ with the exponent about 1/4. This dependence contradicts
to our theory. The reason for discrepancy is not yet clear.
The interval of variation of $h_0$ is not large enough to
establish the exponential dependence reliably, but it definitely
deviates from logarithm. On the other hand, our theory gives 
logarithmic dependence on $h_0$ only if $h_0\gg h_{t1}$. Further
studies both theoretical and experimental are necessary to
clarify the situation.

Our numerical simulations show visibly the formation and propagation
of domain walls as the dominant process at high temperature. The domain
walls look rather rough. It can happen as a consequence of strong
critical fluctuations which are not taken into account in our
DW motion theory, or it may follow from the fact that the dimensionality
2 is the marginal for the development of roughness due to defects
(the roughness exponent $\zeta$ equal to 1 for $d=2$). Nevertheless,
the scaling low ${\cal A}\propto \omega^{1/2}h_0^{1/2}$ for $h\gg h_p$
is confirmed in numerical simulations at high temperatures. 
At lower temperature the nucleation becomes dominant. 

An interesting possibility to verify our predictions 
for the  domain growth controlled hysteresis is a study of 
a hysteresis of adsorption isotherms in a  
close to equilibrium conditions. The process of 
close to equilibrium adsorption can be  studied
experimentally for noble gases (see e.g. \cite{Bruch}). 
The quasiequilibrium desorption of helium films
has been studied in \cite{mw}.
In the case of equilibrium adsorption the chemical potential
plays the role of magnetic field and the coverage - plays the role 
of magnetization. 
The domain wall width for adsorbed systems 
is only one lattice period wide.
For this reason one can expect
rather small nucleation barriers  for equilibrium 
adsorption and 
more broad range of existence 
for domain growth controlled hysteresis
than in magnetic films. More detailed analysis
is given in \cite{lpe}. Though the adsorption
isotherms have been measured starting from Langmuir,
no hysteresis loop measurements has been performed 
until recently. The first  such an experiment 
by  H. Pfn\"ur and K. Budde \cite{pb} gave values
of $\beta $ close to 1/2.

\section{Acknowledgments}
This work was partly supported by the grant of
DOE DE-FG03-96ER 45598 and by the grant of NSF DMR-97-05182.
T.N. acknowledges the support of the German-Israeli Foundation (GIF) and 
the Volkswagen-Foundation and is grateful for the hospitality of the ENS
Paris, where part of this work was done.

We thanks J. Erskine for interesting and fruitful discussions,
and for H.Pfn\"ur for sending us the experimental
results \cite{pb} prior publication.  
One of the authors (V.P.) is grateful to the Sonderforschungsbereich 341
for supporting him during his stay at Cologne University, where this work
has been completed. He also thanks Prof. J. Zittartz for the hospitality
extended to him at Cologne.

\narrowtext
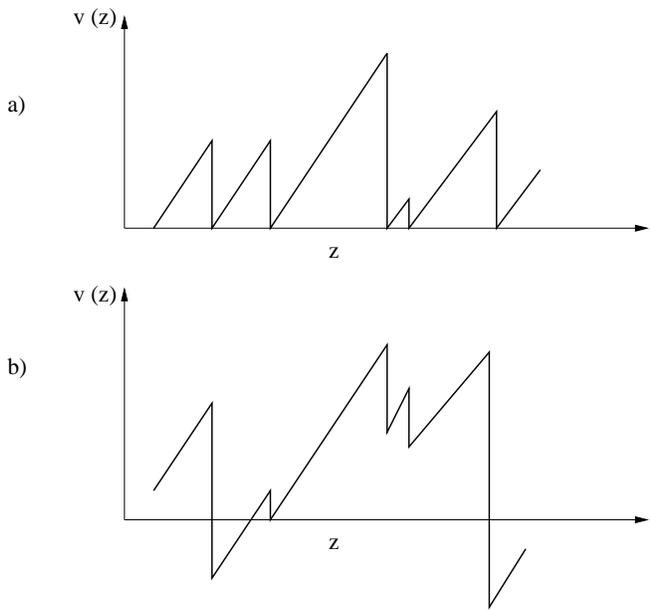
\begin{figure}
\caption{Schematic picture of the ratcheted potential 
in the random bond (a) and random field (b) case.
\label{nat1}}
\end{figure}
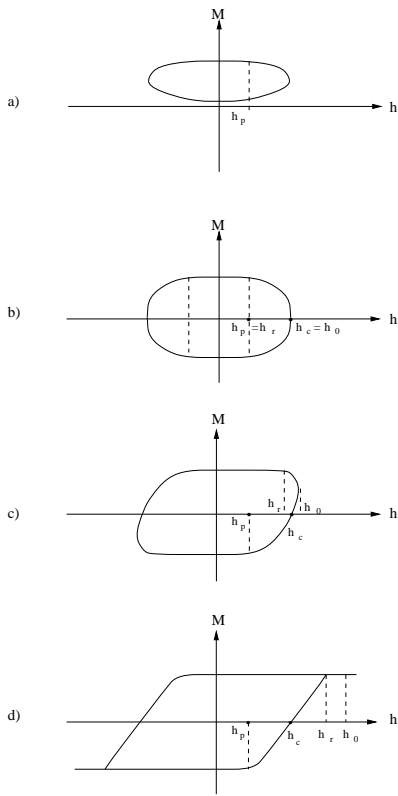
\begin{figure}
\caption{Schematic pictures of hysteresis loops.\\
(a) Incomplete HL for $h_0<h_{t1}$.\\
(b) Symmetric HL for $h_0=h_{t1}$.\\
(c) The HL for $h_{t1}<h_0<h_{t2}$.\\
(d) The HL for $h_0>h_{t2}$.\\
The values $h_p,\; h_c,\; h_r,\; h_0$ are marked in all figures.
\label{fig:loop}}
\end{figure}
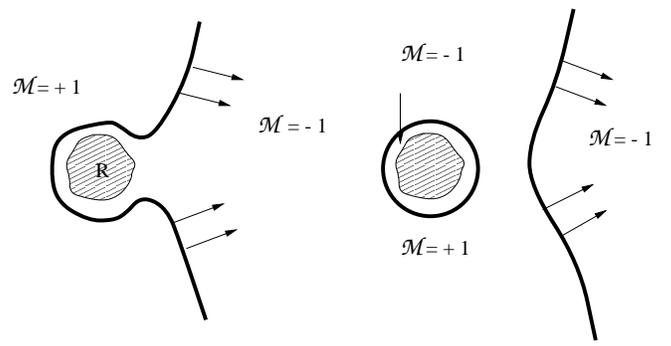
\begin{figure}
\caption{Formation of a bubble by the moving domain wall. 
\label{nat2}}
\end{figure}
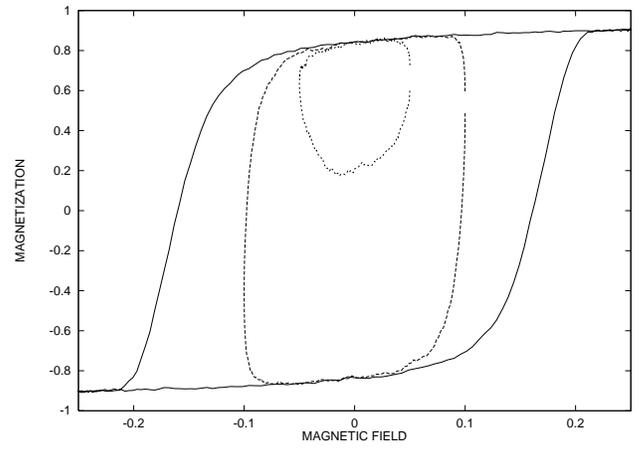
\begin{figure}
\caption{ Simulated hysteresis loops for different values of $h_0$
and constant $\omega$.
The observed hysteresis loop behavior
obey the general classification scheme developed
in Secs.\ref{dw}, \ref{line}  and shown in Fig. \ref{fig:loop}
(see explanation in text).  
\label{f2}}
\end{figure}
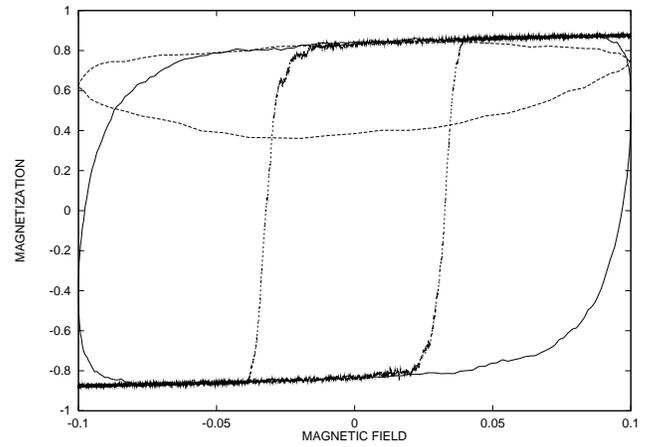
\begin{figure}
\caption{Simulated hysteresis loops for different values of $\omega$
and constant $h_0$.
The observed hysteresis loop behavior
obey the general classification scheme developed
in Secs.\ref{dw},\ref{line}  and shown in Fig. \ref{fig:loop}
(see explanation in text).  
\label{f2a}}
\end{figure}
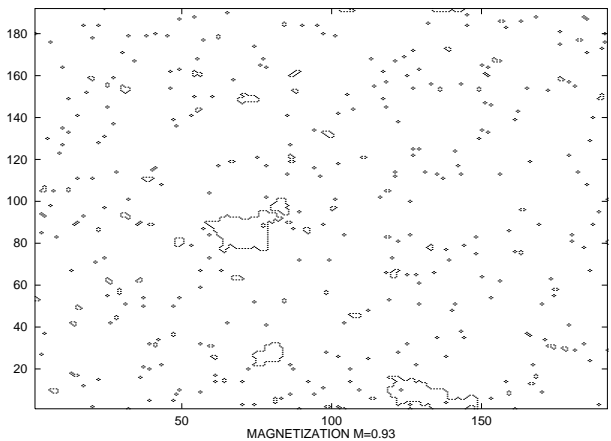
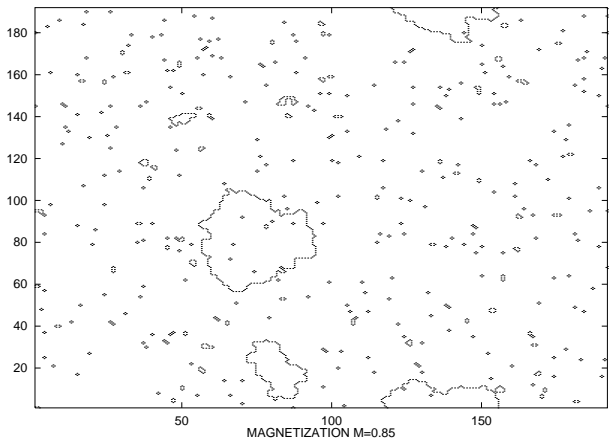
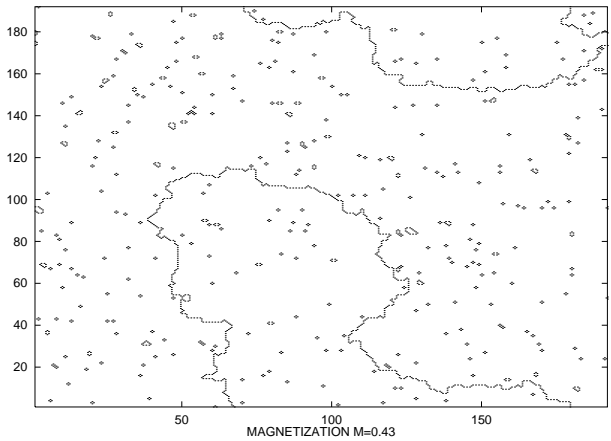
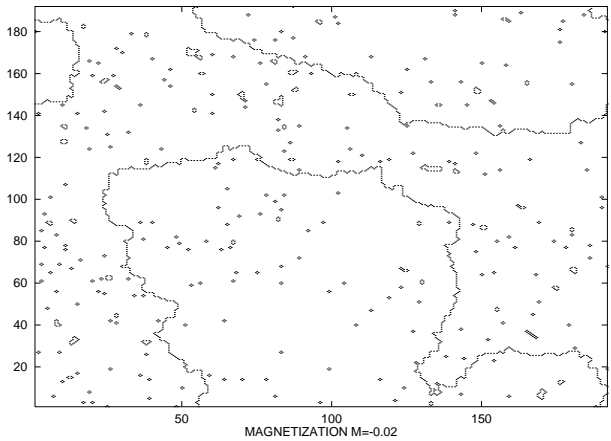
\begin{figure}
\caption{
Inflation of domains with opposite magnetization in an 
Ising model in reversed magnetic field 
with $h_0=0.25 J$ at $T = 0.5 T_c$
for the system of the size $192 \times 192$ with periodic
boundary conditions and with random frozen defects 
(antiferromagnetic bonds) with concentration $0.03$. 
Each dot corresponds to the bond
with opposite spin orientation. 
\label{f6}}
\end{figure}
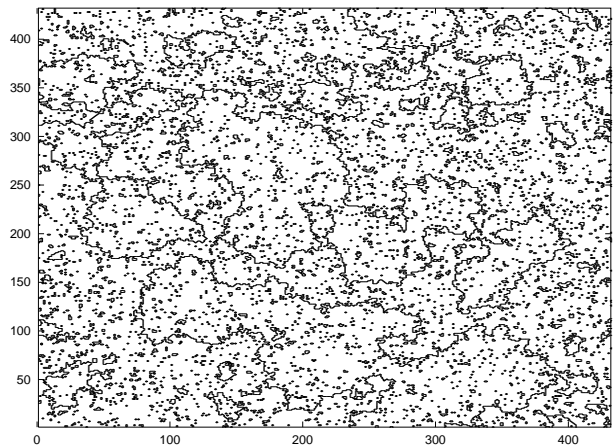
\begin{figure}
\caption{
Domain walls distribution in an 
Ising model in reversed magnetic field with $h_0=0.05 J$ at $T = 0.95 T_c$
for the system of the size $432 \times 432$ with periodic
boundary conditions. Each dot corresponds to the bond
with opposite spin orientation.  
\label{f1}}
\end{figure}
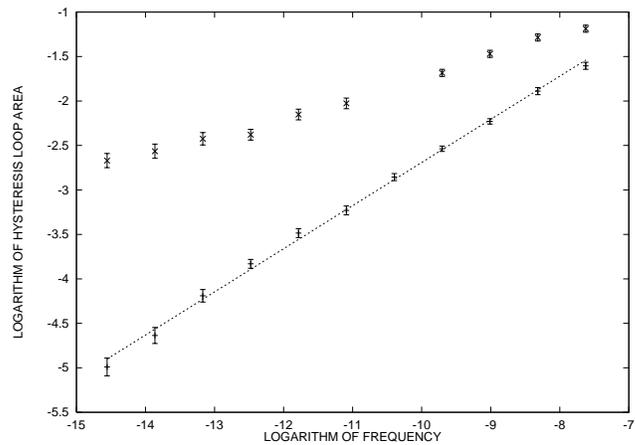
\begin{figure}
\caption{ Hysteresis loop area as a function
of frequency in a double logarithmic coordinates
for the magnetic field  with amplitude $h_0 =0.05 J$ 
at $T = T_c$ (lower line) and  at $T = 0.95 T_c$.
\label{f7}}
\end{figure}
\end{multicols}

\begin{references}
\bibitem{footnote}
{Sue and Erskine  claimed a power dependence with
a small exponent, but in the frameworks of their data it is
indistinguishable from the logarithm.}
\end{references}
\end{document}